# Spectroscopic confirmation of four metal-poor galaxies at z=10.3-13.2

Emma Curtis-Lake[1*], Stefano Carniani[2†], Alex Cameron[3], Stephane Charlot[4], Peter Jakobsen[5,6], Roberto Maiolino[7,8,9], Andrew Bunker[3], Joris Witstok[7,8], Renske Smit[10], Jacopo Chevallard[3], Chris Willott[11], Pierre Ferruit[12], Santiago Arribas[13], Nina Bonaventura[5,6], Mirko Curti[7,8], Francesco D'Eugenio[7,8], Marijn Franx[14], Giovanna Giardino[15], Tobias J. Looser[7,8], Nora Lützgendorf[16], Michael V. Maseda[17], Tim Rawle[16], Hans-Walter Rix[18], Bruno Rodríguez del Pino[13], Hannah Übler[7,8], Marco Sirianni[16], Alan Dressler[19], Eiichi Egami[20], Daniel J. Eisenstein[21], Ryan Endsley[22], Kevin Hainline[20], Ryan Hausen[23], Benjamin D. Johnson[21], Marcia Rieke[20], Brant Robertson[24], Irene Shivaei[20], Daniel P. Stark[20], Sandro Tacchella[7,8], Christina C. Williams[25], Christopher N. A. Willmer[20], Rachana Bhatawdekar[26], Rebecca Bowler[27], Kristan Boyett[28,29], Zuyi Chen[20], Anna de Graaff[18], Jakob M. Helton[20], Raphael E. Hviding[20], Gareth C. Jones[3], Nimisha Kumari[30], Jianwei Lyu[20], Erica Nelson[31], Michele Perna[13], Lester Sandles[7,8], Aayush Saxena[3,9], Katherine A. Suess[24,32], Fengwu Sun[20], Michael W. Topping[20], Imaan E. B. Wallace[3] and Lily Whitler[20]

[1]Centre for Astrophysics Research, Department of Physics, Astronomy and Mathematics, University of Hertfordshire, Hatfield AL10 9AB, UK. [2]Scuola Normale Superiore, Piazza dei Cavalieri 7, I-56126 Pisa, Italy. [3]Department of Physics, University of Oxford, Denys Wilkinson Building, Keble Road, Oxford OX1 3RH, UK. [4]Sorbonne Université, CNRS, UMR 7095, Institut d'Astrophysique de Paris, 98 bis bd Arago, 75014 Paris, France. [5]Cosmic Dawn Center (DAWN), Copenhagen, Denmark. [6]Niels Bohr Institute, University of Copenhagen, Jagtvej 128, DK-2200, Copenhagen, Denmark. [7]Kavli Institute for Cosmology, University of Cambridge, Madingley Road, Cambridge, CB3 0HA, UK. [8]Cavendish Laboratory - Astrophysics Group, University of Cambridge, 19 JJ Thomson Avenue, Cambridge, CB3 0HE, UK. [9]Department of Physics and Astronomy, University College London, Gower Street, London WC1E 6BT, UK. [10]Astrophysics Research Institute, Liverpool John Moores University, 146 Brownlow Hill, Liverpool L3 5RF, UK. [11]NRC Herzberg, 5071 West Saanich Rd, Victoria, BC V9E 2E7, Canada. [12]European Space Agency, European Space Astronomy Centre, Madrid, Spain. [13]Centro de Astrobiología (CAB), CSIC–INTA, Cra. de Ajalvir Km. 4, 28850- Torrejón de Ardoz, Madrid, Spain. [14]Leiden Observatory, Leiden University, P.O. Box 9513, NL-2300 RA Leiden, Netherlands. [15]ATG Europe for the European Space Agency, ESTEC, Noordwijk, The Netherlands. [16]European Space Agency (ESA), ESA Office, STScI, Baltimore, MD 21218, USA. [17]Department of Astronomy, University of Wisconsin-Madison, 475 N. Charter St., Madison, WI 53706, USA. [18]Max-Planck-Institut für Astronomie, Königstuhl 17, D-69117, Heidelberg, Germany. [19]The Observatories of the Carnegie Institution for Science, 813 Santa Barbara St., Pasadena, CA 91101, USA. [20]Steward Observatory University of Arizona 933 N. Cherry Avenue ,Tucson, AZ 85721, USA. [21]Center for Astrophysics, Harvard & Smithsonian, 60 Garden St., Cambridge, MA 02138, USA. [22]Department of Astronomy, University of Texas, Austin, TX 78712, USA. [23]Department of Physics and Astronomy, The Johns Hopkins University, 3400 N. Charles St., Baltimore, MD 21218, USA. [24]Department of Astronomy and Astrophysics University of California, Santa Cruz, 1156 High Street, Santa Cruz, CA 96054, USA. [25]NSF's National Optical-Infrared Astronomy Research Laboratory, 950 North Cherry Avenue, Tucson, AZ 85719, USA. [26]European Space Agency, ESA/ESTEC, Keplerlaan 1, 2201 AZ Noordwijk, NL. [27]Jodrell Bank Centre for Astrophysics, Department of Physics and Astronomy, School of Natural Sciences, The University of Manchester, Manchester, M13 9PL, UK. [28]School of Physics, University of Melbourne, Parkville 3010, VIC, Australia. [29]ARC Centre of Excellence for All Sky Astrophysics in 3 Dimensions (ASTRO 3D), Australia. [30]AURA for European Space Agency, Space Telescope Science Institute, 3700 San Martin Drive, Baltimore, MD 21218, USA. [31]Department for Astrophysical and Planetary Science, University of Colorado, Boulder, CO 80309, USA. [32]Kavli Institute for Particle Astrophysics and Cosmology and Department of Physics, Stanford University, Stanford, CA 94305, USA.

[*] Corresponding author(s). E-mail(s): e.curtis-lake@herts.ac.uk; stefano.carniani@sns.it;

[†]These authors contributed equally to this work.

**Finding and characterising the first galaxies that illuminated the early Universe at cosmic dawn is pivotal to understand the physical conditions and the processes that led to the formation of the first stars. In the first few months of operations, imaging from the James Webb Space Telescope (JWST) has been used to identify tens of candidates of galaxies at redshift ($z$) greater than 10 less than 450 million years after the Big Bang. However, none of such candidates has yet been confirmed spectroscopically, leaving open the possibility that they are actually low-redshift interlopers. Here we present spectroscopic confirmation and analysis of four galaxies unambiguously detected at redshift 10.3≤z≤13.2, previously selected from NIRCam imaging. The spectra reveal that these primeval galaxies are metal poor, have masses between of order ~$10^7$-$10^8$ solar masses, and young ages. The damping wings that shape the continuum close to the Lyman edge provide the first constraints on neutral Hydrogen fraction of the intergalactic medium to be obtained from normal star-forming galaxies. These findings demonstrate the rapid emergence of the first generations of galaxies at cosmic dawn.**

The opening act of galaxy formation in the first billion years after the Big Bang sets in motion the physics of galaxy formation and evolution that shapes galaxy properties across cosmic time. Galaxies forming at these times may be the seeds of the much more massive and mature galaxies in the local Universe. Theoretical models and cosmological simulations differ greatly in their predictions of the physical properties and abundance of the first galaxies. The theoretical pictures depend strongly on assumptions about the physical processes at play in the early universe, such as: gas cooling and fragmentation in primordial clouds; the feedback effects from first stars and supernova explosions that subsequently enrich the surrounding medium; and early merging, assembly and accretion histories of galaxies[1-7]. The abundance and mass distribution of the first galaxies are also tightly connected to early structure formation. Therefore, the detection and characterisation of these early galaxies is key to test different models and theories.

High-redshift galaxies often have distinctive spectra in the ultraviolet, in which the blue spectrum produced by hot massive stars is abruptly cut off below the Lyman-limit at 912Å (rest-frame) by the absorption of the light by neutral Hydrogen in stellar atmospheres, interstellar gas and the intergalactic medium (IGM). At the highest redshifts ($z \gtrsim 6$), the intergalactic neutral Hydrogen leads to almost complete absorption at wavelengths below Ly$\alpha$ at 1216Å. Observationally, this translates to a 'dropout', i.e., a lack of detection in bands blue-ward of $(1 + z) \times 1216$ Å but flux red-ward of the same wavelength[8-10]. However, galaxies with peculiar properties may mimic high-redshift galaxies [e.g., a combination of dust reddening and nebular lines or contribution by an active galactic nucleus as in ref 11]. Therefore, spectroscopic observations are the only method to determine accurate redshifts, either via the detection of the (redshifted) nebular lines[12], or via the unambiguous detection of the sharp continuum cutoff at $(1 + z) \times 1216$ Å. The highest redshift spectroscopically confirmed galaxy prior to these observations is that of GN-z11 at z=10.957[13,14].

Identification and spectroscopic characterisation of galaxies in the early Universe is one of the primary goals for which JWST was designed. The first few months of JWST imaging have already yielded a large number of candidate galaxies at $z > 10$[15-22]. However, the redshift estimates of these candidates have so far been based on their broad-band spectral energy distributions (SEDs), and it cannot be ruled out that such candidates are actually lower redshift galaxies[11], especially in regions where accompanying Hubble space telescope imaging is relatively shallow. With the large number of z>10 candidates identified in the first months of JWST science observations from NIRCam photometry, some initial findings suggest very little evolution of the UV luminosity function above z>10[18,23] (though this is not seen in ref 15). This would require early galaxies to display different physics, for example a stellar initial mass function more top-heavy than in lower-redshift galaxies[18]. Yet ref. 24 illustrates how large a difference in UV luminosity density evolution is measured when considering only the robust candidates. This demonstrates the firm need for spectroscopic observations to follow up photometric candidates. Additionally, spectra provide us with constraints on the stellar and gas properties of the objects beyond what photometry can provide.

We report here the deepest spectroscopic observations to date with NIRSpec[25] on JWST, which provide confirmation of four candidates at z>10 and extensive characterisation of their physical properties. These candidates were photometrically identified as part of the JWST Advanced Deep Extragalactic Survey (JADES), a joint guaranteed time project of the NIRCam and NIRSpec instrument teams. The identification and photometric study of these candidates, based on Hubble Space Telescope (HST) and NIRCam data[26], is described in a companion paper[27]. We specifically focus here on a pointing in the Hubble Ultra Deep Field (in the GOODS-South area), in which we have taken multi-object spectroscopy of 253 galaxies observed simultaneously with NIRSpec's configurable array of microshutters. We report here on observations taken with the prism spectral configuration (spectral range 0.6–5.3μm, resolving power R ∼ 100) with exposure times ranging from 9.3 to 28 hours (see Methods for details on the observing strategy).

The JADES spectroscopic observations reach an unprecedented sensitivity of 28.4 magnitudes (AB) at 5σ per resolution element on the continuum at 2.5μm. We note that the NIRSpec prism is extremely well-suited for the redshift confirmation of high-z candidates, with low spectral resolution and high sensitivity at short wavelengths where we are searching for a spectral break (around 1–2μm), and higher resolution in the 3–5μm region, where we are searching for narrow spectral lines.

The focus of this paper is on four of these spectroscopic targets. Two of these are z > 12 galaxy candidates selected from NIRCam imaging[27], based on a clear lack of F150W flux. Two others are z > 10 candidates based on their HST IR photometry. We defer to a future publication to describe the other targets in this deep pointing. All candidates are faint, with F200W magnitudes fainter than 28 (AB), and hence entirely out of reach for any spectroscopic facility before JWST. More details on the selection and photometric properties of these candidates are provided in the companion paper[27].

In Figure 1 we show the 1D and 2D spectra of these four galaxies. All show a clear detection of a blue continuum that drops off sharply in a manner consistent with a z > 10 Lyman dropout. Specifically, in Figure 1, we show the redshift derived from the position of the spectral break, taken to be at the wavelength of Ly$\alpha$ at

1215.67Å. These redshifts are reported in Table 1 and were derived with full spectral fitting over the entire redshift range, with each object consistently showing peaks in the posterior probability distribution only at high redshifts. We note that both the spectra and the photometry from ref. 27 agree on the wavelength of the break.

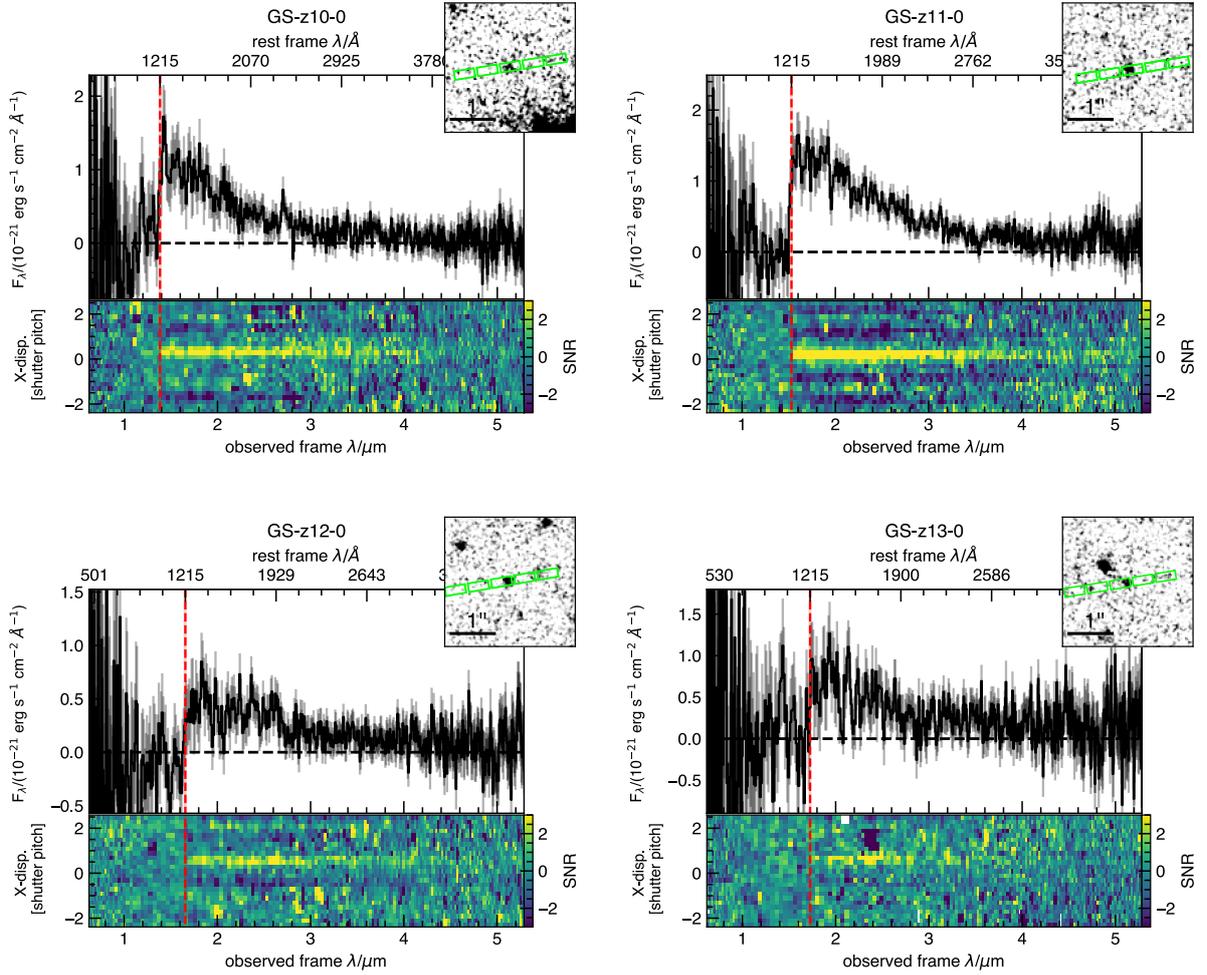

**Fig. 1** NIRSpec prism R ~ 100 spectra for the four z > 10 galaxies targeted for the first deep spectroscopic pointing of the JADES survey, JADES-GS-z10-0, JADES-GS-z11-0, JADES-GS-z12-0 and JADES-GS-z13-0. For each galaxy we display the 1D spectrum and associated $1\sigma$ uncertainties (which are derived from standard error propagation through the reduction pipeline). In the bottom panel we show the 2D signal-to-noise ratio plot. The 2D plot is binned over four pixels in the wavelength direction to better show the contrast across the break. The inset panel in the top right-hand corner shows the NIRCam F444W filter image with the three nodding positions of the NIRSpec micro-shutter 3-slitlet array aperture shown in green. The red dashed line shows 1215.67Å at the observed redshift $z_{1216}$.

We tested whether the observed breaks might be produced by the Balmer (or 4000Å) break in the stellar continuum associated with evolved stellar populations in galaxies at z ~ 3, and we show that this possibility is excluded with high confidence (see Methods section and model Balmer Break strength in Extended Data Figure 1). Other, more extreme low redshift solutions are still able to explain the photometry of our objects. In particular, fitting with active-galactic-nuclei narrow-line emission (not often included when performing photometric redshifts) provides intermediate redshift solutions at z~3-3.5, as shown in the Methods section. In these solutions the strong photometric break is produced by strong line emission in certain filters. We can firmly rule out these low-redshift solutions with our spectra, which demonstrates that the photometric fluxes arise from continuum emission, rather than strong emission lines with weak underlying continuum. Although this low redshift solution may appear extreme, we emphasise that JWST is pushing into a new regime of exploration, where yet unexplored families of low-redshift contaminants may be uncovered. A recent discovery of a triply-imaged point source shows a similarly extreme possible SED explained by strong line emission at $z_{phot}$7.7[28].

One of the galaxies, JADES-GS-z11-0, has been debated in the literature. It was first identified in ref. 29 as a potential z ~ 10 galaxy. In the UDF12 survey[30,31], deep Hubble imaging revealed the object drops out in $JH_{140}$ band imaging. This left two possibilities, either the source was at very high redshift [z ~ 11.9[30]], or was low redshift, with a high-equivalent-width emission line producing the flux in the Hubble $H_{160}$ image. In fact, ref. 32 found indication of a possible emission line in spectroscopic follow-up supporting the latter explanation. We do not confirm this line emission in our NIRSpec spectroscopy, and the NIRCam imaging present in ref. 27 shows that the continuum emission extends to longer wavelengths, which is consistent with the spectrum shown here [see also ref. 23]. Therefore, this galaxy is indeed at high redshift, and not a low-redshift contaminant.

The continua appear mostly featureless, with the possible exceptions of JADES-GS-z10-0, which shows a tentative emission line at ~ 1.44 μm which may indicate Lyα emission, and JADES-GS-z12-0, which shows another tentative feature at ~ 5.23 μm, which could be interpreted as [Ne III]λ3869 emission at z = 12.52. However, both features are only marginally detected, and we are still assessing whether these very faint and localized features are astrophysical. They will be assessed and explored more in detail in forthcoming papers and not discussed further here. Remarkably, the lack of strong line detections turns out to be what makes these spectra particularly interesting as it provides vital information over and above the spectroscopic redshift determination, and what could be derived from photometry alone[27].

**Table 1** Exposure times, redshifts (derived both from assuming the spectral break is at exactly 1215.67Å and accounting for the damping wing from a fully neutral IGM), 2σ upper limits on emission-line equivalent widths (rest frame) for the C III]λλ1907,1909 He IIλ1640 and [O II]λλ3726,3729 lines, 2σ lower limits on the strength of the observed spectral breaks (measurements described in Methods 2), UV absolute magnitude, $M_{uv}$, and UV slope, β (measured directly from the spectra, see Methods 2.3) and BEAGLE-derived physical properties for the four objects. For the BEAGLE-derived properties we report posterior medians and limits in the 1σ credible region.

| JADES-ID | GS-z10-0 | GS-z11-0 | GS-z12-0 | GS-z13-0 |
|---|---|---|---|---|
| Full name | JADES-GS+53.15884-27.77349 | JADES-GS+53.16476-27.77463 | JADES-GS+53.16634-27.82156 | JADES-GS+53.14988-27.77650 |
| Exposure time (s) | 67225.6 | 100838.0 | 67225.6 | 33612.8 |
| $z_{1216}$ * | $10.38^{+0.07}_{-0.06}$ | $11.58^{+0.05}_{-0.05}$ | $12.63^{+0.24}_{-0.08}$ | $13.20^{+0.04}_{-0.07}$ |
| $z_{HI}$ † | $10.37^{+0.03}_{+0.02}$ | $11.48^{+0.03}_{-0.08}$ | $12.6^{+0.04}_{-0.05}$ | $13.17^{+0.16}_{-0.15}$ |
| EW(C III]) Å 2σ | < 13.8 | < 5.9 | < 12.4 | < 15.2 |
| EW(He II)/Å 2σ | < 14.8 | < 6.0 | < 13.5 | < 15.4 |
| EW(O II)/Å 2σ | < 28.1 | < 9.1 | < 16.6 | < 16.8 |
| 2σ break strength | > 2.04 | > 6.85 | > 2.48 | > 2.79 |
| $M_{UV}$ | −18.61 ± 0.10 | −19.34 ± 0.05 | −18.23 ± 0.16 | −18.73 ± 0.06‡ |
| β | −2.49 ± 0.22 | −2.18 ± 0.09 | −1.84 ± 0.19 | −2.37 ± 0.12‡ |
| $log(M/M_\odot)$ | $7.58^{+0.19}_{-0.20}$ | $8.67^{+0.08}_{-0.13}$ | $7.64^{+0.66}_{-0.39}$ | $7.95^{+0.19}_{-0.29}$ |
| $\Psi / M\ yr^{-1}$ ¶ | $1.1^{+0.19}_{-0.16}$ | $2.2^{+0.28}_{-0.22}$ | $1.8^{+0.54}_{-0.43}$ | $1.36^{+0.31}_{-0.23}$ |
| $log(t/yr)$ ‖ | $7.54^{+0.25}_{-0.20}$ | $8.35^{+0.08}_{-0.17}$ | $7.36^{+0.75}_{-0.59}$ | $7.84^{+0.23}_{-0.36}$ |
| $log(Z/Z_\odot)$ ¤ | $-1.91^{+0.25}_{-0.20}$ | $-1.87^{+0.28}_{-0.18}$ | $-1.44^{+0.23}_{-0.22}$ | $-1.69^{+0.28}_{-0.31}$ |
| $\hat{\tau}_v$ †† | $0.05^{+0.03}_{-0.02}$ | $0.18^{+0.06}_{-0.06}$ | $0.17^{+0.20}_{-0.09}$ | $0.10^{+0.08}_{-0.05}$ |
| $\xi_{ion}$ § | $25.46^{+0.07}_{-0.07}$ | $25.43^{+0.06}_{-0.06}$ | $25.72^{+0.16}_{-0.19}$ | $25.47^{+0.10}_{-0.09}$ |

*The redshift based on the spectral break being at 1215.67Å †The redshift accounting for a fully neutral IGM ($x_{HI}$ = 1) following the method outlined in Methods 2.1. ‡For JADES-GS-z13-0, we report β and $M_{uv}$ derived from the beagle fitting, since we know this object to be on the edge of the shutter, and hence incorporate NIRCam photometry in the fitting to this one object to account for slit-losses (see Methods 3). ¶Ψ is the star formation rate. ‖t is age of the oldest stars, or maximum stellar age. ¤Z is the metallicity. ††$\hat{\tau}_v$ is the effective V-band attenuation optical depth. §The production rate of H-ionizing photons per unit monochromatic UV luminosity.

Leaving aside the above two features, we report the 2σ upper limits on the equivalent widths (EW) of He IIλ1640, C III]λλ1907,1909 and [O II] λλ3726,3729 in Table 1 (see Methods 2.2 for details of the measurements and Extended Data Table 1 for further limiting fluxes). These lines are important as C III]λλ1907,1909 is often the strongest line in the rest-frame UV of low-metallicity galaxies [e.g. ref. 33] and strong He IIλ1640 is expected in galaxies of near-zero metallicity. Nearby metal-poor galaxies ($Z/Z_\odot \lesssim 0.1$) show EWs of C III]λλ1907, 1909 spanning ~ 6–16Å[33,34], meaning our limits are constraining only in the upper region of this range. The limits on EW(He IIλ1640) are above the actually measured equivalent widths in these studies, and comparable or higher than the majority of objects in the MUSE-selected sample of ref. 35 spanning $2 \lesssim z \lesssim 4$. However, JADES-GS-z10-0 and JADES-GS-z11-0 still show strong limits on the equivalent width of [O II]λλ3726,3729. These limits are constraining, showing that [O II]λλ3726,3729 is weak compared to the average equivalent widths measured in z~3 galaxies (~80Å at $10^9$ solar masses)[36].

We can gain further insight into the physical properties of these galaxies through spectral fitting with the BEAGLE [BayEsian Analysis of GaLaxy sEds[37]] tool with setup described in Extended Data Table 2, adopting a constant star formation history (SFH) to probe the young stellar populations within them (see Methods section 3 for more details). These fits are illustrated in Extended Data Figures 2–5. We find low metallicities, with two galaxies, JADES-GS-z10-0 and JADES-GS-z11-0, showing strong constraints of just a few percent of solar. At higher metallicities we would expect a significant detection of [OII]λλ3726,3729. For the two highest redshift galaxies, a low metallicity is still preferred, but the constraints are less strong due to [O II]λλ3726,3729 being very close to the edge of the observable wavelength coverage of NIRSpec, where the S/N is low, as well as lower S/N in regions of the rest-frame UV lines.

The measured star formation rates are moderate, at just a few solar masses a year. We caution that total star formation rates and stellar masses require slit-loss corrections which can be best derived from NIRCam photometry if the objects are extended, or on the edge of the shutter. Still, we find excellent agreement in absolute magnitude and UV slope compared to those derived from NIRCam photometry alone[27] for the two galaxies well-centred within the microshutters (GS-z10-0 and GS-z11-0). The fitted parameters are given in Table 1.

It is of interest to investigate whether the lack of detectable line emission requires a large escape fraction of ionising photons and find no strong dependence of the physical properties, nor strong constraints on the escape fraction. We do note, however, that JADES-GS-z11-0 has a solution with low ages and high escape fraction with marginally higher metallicity, though the upper 1σ limit is still ∼ 5% solar metallicity (see Extended Data Table 3). In fact, at these extremely low metallicities, the rest-frame UV emission lines might be significantly weaker than those sometimes found in lower redshift samples with somewhat higher metallicities around ∼ 10% solar. Another important factor is the age of the stellar populations, as strong UV emission lines have primarily been observed in galaxies with UV light dominated by very young stellar populations [≲ 10 million years[33]].

Two objects, JADES-GS-z11-0 and JADES-GS-z12-0 do indicate moderate levels of dust, albeit with large uncertainties (with effective V-band absorption optical depth, $\hat{\tau}_v = 0.18^{+0.06}_{-0.06}$ and $\hat{\tau}_v = 0.2^{+0.16}_{-0.09}$, respectively). Such high optical depth due to dust in such low metallicity systems would be physically hard to explain unless a low dust-to-gas ratio were integrated over large HI column densities. However, we note that some recent models expect significant dust production by the first generations of stars[3,38]. When fit without any dust, JADES-GS-z11-0 required a higher metallicity (up to 40% solar at 1σ), while the JADES-GS-z12-0 data yielded similar parameter constraints to those reported in Table 1.

The spectra do not provide strong constraints red-ward of the Balmer-break, the longest rest-frame wavelength probed being ∼ 3660 – 4350Å. We see no evidence for strong Balmer breaks in these objects, but the S/N in this region of the spectra are low. Correspondingly, the constraints on stellar age and stellar mass (sensitive to the Balmer-break strength) are broad. The ages range from ∼ 5 to 230 Myr, and stellar masses range from ∼ 2 × $10^6$ M$_\odot$ to 460×$10^6$ M$_\odot$, though the constraints on these parameters are weak, and highly sensitive to the prior regarding the time history of the star formation rate. The associated production rates of H-ionizing photons per unit monochromatic UV luminosity, $\xi_{ion}$, are similar to those measured in extreme star-forming regions in low-redshift, metal-poor galaxies[39].

We show the measured UV slopes (β) vs. absolute magnitude at 1500Å (M$_{uv}$) in Figure 2. We compare to other JWST-selected high-redshift candidate samples spanning photometric redshifts z ∼ 7 – 16 [22,40,41]. Our measured slopes at such faint M$_{uv}$ magnitudes are comparable to the other literature samples, suggesting little evolution at these epochs. Very blue UV slopes are used to search for extreme stellar populations at the earliest times[42]. In this case, we find extreme stellar populations but the presence of any nebular recombination continuum will redden the UV slopes[41]. Indeed, we expect strong nebular continuum emission in low metallicity galaxies unless a significant fraction of their ionising photons escape into the intergalactic medium.

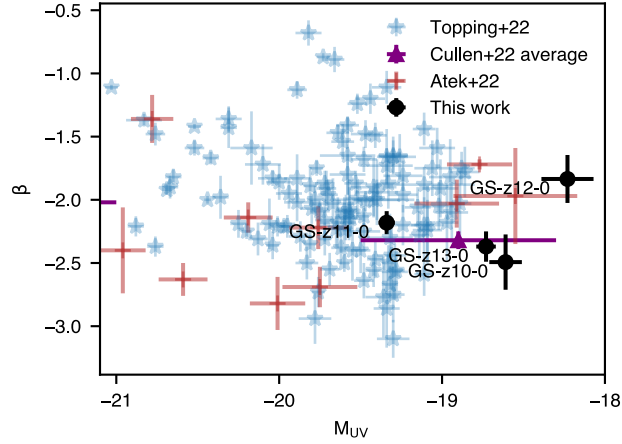

**Fig. 2** UV slope, β, as a function of absolute magnitude at 1500Å, $M_{UV}$, measured as described in Methods 2.3. These are compared to the measurements from photometrically selected high-redshift candidates. Specifically, the average β measured from objects spanning z ∼ 8 – 15 at similar $M_{UV}$ from Cullen+22[40], as well as a sample from Atek+22[22] spanning z ∼ 9 – 16, and the sample presented in Topping+22[41] (itself collated from the samples of refs. 21 and 43 and spanning z ∼ 7 – 11). In all cases the error bars show the 1σ measurement uncertainties, except in the case of the point showing the average UV slope from Cullen+22. For the Cullen+22 datapoint, the point shows the inverse-variance weighted mean and standard error of β, plotted against the median $M_{UV}$ of their sample of 41 galaxies in their lower luminosity bin. The errorbar in $M_{UV}$ is $\sigma_{MAD}$ (where $\sigma_{MAD} = 1.483\ x\ MAD$ and MAD is the median absolute deviation) of the individual $M_{UV}$ values.

If ionising radiation does escape from galaxies, it will reionize neutral Hydrogen in the surrounding gas. We find that the breaks in the spectra presented in Fig. 1 are significantly less abrupt than those seen in galaxies at lower redshifts in our spectroscopic data set, and are consistent with a softening of the break by the Lyα damping wing caused by a largely neutral IGM, suggesting that these galaxies are yet to ionize large bubbles in their near vicinity. The redshift is sensitive to the existence and form of this damping wing, and we report the best-fit redshift for a fully neutral inter-galactic medium ($x_{HI} = 1$, where $x_{HI}$ is the fraction of neutral Hydrogen) in Table 1. For the object with the highest S/N in the Lyα break region (JADES-GS-z11-0), we investigate the constraints we can place on the neutral Hydrogen fraction by first fixing the best-fit model for the stellar population. We then run BEAGLE varying only redshift and $x_{HI}$. The resulting fit to the spectral break, and the derived constraints are shown in Figure 3. The 2D posterior probability distribution function reported in the right-hand panel indicates that the constraints on $x_{HI}$ are fairly weak, suggesting $x_{HI}>0.5$ from the 1σ credible interval, though this is sensitive to exact redshift of the source (e.g. Extended Data Figure 6). However, these spectra demonstrate that $x_{HI}$ can be constrained from JWST R100 spectra at slightly lower redshifts from 'normal' star-forming galaxies.

To date the only constraints on the evolution of $x_{HI}$ from damping wings is in luminous quasars at 6 < z < 7.5. Damping wings are rarely observed at z < 7, but become more common in the small sample of known z > 7 quasars, consistent with $x_{HI}$ ∼ 0.5 at z = 7.3[44]. Whilst star-forming galaxies are less luminous than quasars, they have several advantages in being plentiful at redshifts 7 to 9 and providing an independent test of neutrality at higher redshifts. Finally, galaxies do not exhibit broad Lyα and N vλ1240 emission, which may simplify the damping wing modelling[45].

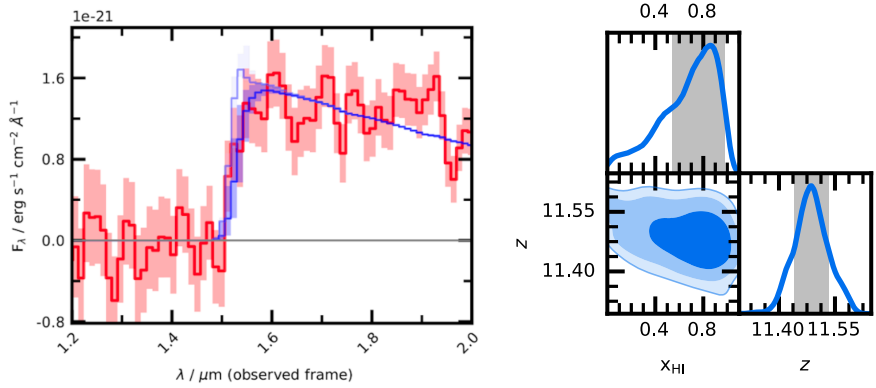

**Fig. 3** BEAGLE fit to the spectral break region of JADES-GS-z11-0 (top panel), while varying the fraction of neutral Hydrogen in the IGM, $x_{HI}$. The red line and shaded region show the extracted spectrum and per-pixel $1\sigma$ uncertainties and the darker blue line and shaded region shows the range and median of the fitted models, respectively. The lighter blue line shows the underlying intrinsic spectrum before application of the damping wing. The bottom panel shows the 2D constraints on redshift and $x_{HI}$, which were varied in the fit while keeping all other physical properties constant (see text for details). For this test, we use a spectral extraction over 3 pixels that maximises the S/N in the region of the break. The shape of the damping wing is not sensitive to wavelength-dependent slit losses introduced by such a small extraction box since the wing it extends over just tens of pixels.

We conclude by emphasizing that is the results reported here represent a milestone for the JWST mission, pushing the spectroscopic frontier to a markedly earlier epoch of galaxy formation. In addition to providing clear detections of Lyman dropouts as high as z = 13.2, these JADES observations also show the power of spectroscopy to probe the physics of these galaxies, revealing low metallicities by through the lack of emission lines, as well as the state of the surrounding intergalactic medium. This is just a starting point for the mission. JADES and other programmes have extensive amounts of spectroscopy approved for JWST-detected high-redshift candidates.

# Methods

## 1. NIRSpec observations and data reduction

The NIRSpec observations presented here are part of GTO program ID: 1210 (Principal Investigator: Lützgendorf) and were obtained between October 22 and 25, 2022. The program used a three-point nod pattern for background subtraction, as well as three small dithers with microshutter array (MSA) reconfigurations in order to improve spatial sampling, increase sensitivity and flux accuracy, mitigate the impact of the detector gaps, and aid removal of cosmic rays.

Each dither pointing included four sequences of three nodded exposures each to build up signal-to-noise. Observations were carried out by using the disperser-filter combination PRISM/CLEAR, which covers the wavelength range between 0.6 μm and 5.3 μm and provides spectra with a spectral power of R ∼ 100 [25]. Each PRISM/CLEAR setup had two integrations of 19 groups, resulting in an exposure time of 8403.2 seconds for each sequence and of 33,612 seconds for each dither pointing.

A total of 253 galaxies were observed over the three dither pointings. As the non-functioning shutters and rigid grid of the MSA prevents some slit locations from being used, some galaxies were not observed on all three pointings. More specifically, among the four sources presented in this paper, JADES-GS-z11-0 was observed in all three MSA dither pointings, JADES-GS-z10-0 and JADES-GS-z12-0 were present in two dither pointings, whereas JADES-GS-z13-0 was only observed in one dither pointing. The different resulting exposure times for each target are reported in Table 1.

Flux-calibrated 2D spectra and 1D spectral extractions have been produced using pipelines developed by the ESA NIRSpec Science Operations Team (SOT) and the NIRSpec GTO Team. We briefly outline here the main steps, while a more detailed description will be presented in a forthcoming NIRSpec/GTO collaboration paper. Most of the processing steps in the pipelines adopt the same algorithms as included in the official STScI pipeline used to generate the MAST archive products [see Fig. 11 and section 4.3 of ref 46]. Initially, we processed the MOS raw data (i.e, level 1a data from the MAST archive) with the ramp-to-slope pipeline which estimates the count rate per pixel by using all unsaturated groups in the ramp. Ramp jumps due to cosmic rays are detected and rejected on the basis of the slope of the individual ramps. The ramp-to-slope pipeline also includes the following steps: saturation detection and flagging, master bias subtraction, reference pixel subtraction, linearity correction, dark subtraction, snowball artifact detection and correction, and count rate estimation [for more details see refs. 47-49]. All the count-rate images were then processed using a data reduction pipeline including ESA NIRSpec SOT codes and NIRSpec GTO algorithms. The pipeline has 11 main steps: 1) identification of non-target galaxies intercepting the open shutters; 2) pixel-level background subtraction by combining the three nod exposures (excluding nods contaminated by non-target sources); 3) extraction of sub-images containing the spectral trace of each target and wavelength and spatial coordinate assignments to each pixel in the 2D maps; 4) pixel-to-pixel flat-field correction; 5) spectrograph optics and dispersers correction; 6) absolute flux calibration; 7) slit-losses correction; 8) rectification of the spectral trace; 9) extraction of 1D spectra; 10) combination of 1D spectra generated from each integration, nod, and pointing; 11) combination of 2D maps. The data processing workflow thus returns both a combined 1D and 2D spectrum for each target. We stress however that the combined 1D spectra are not extracted from the combined 2D maps, but are the result of a weighted average of 1D spectra from all integrations. This process allowed us to mask the bad pixels indicated on the quality flags and to reject outlier pixels. Finally, we adopted an irregular wavelength grid for the 1D and 2D spectra to avoid oversampling of the line spread function at short wavelengths (λ ∼ 1 μm).

Given the compact size of our z > 10 targets, we computed and applied slit-loss corrections, modelling galaxies as point-like sources, but taking into account the relative intra-shutter position of each source (each microshutter has an illuminated area of 0.2"×0.46"). For each target we extracted the 1D spectra from two different apertures. One aperture was as large as the shutter size to recover all emission of the galaxy, while the second extraction was performed in an aperture of 3-pixels height (with NIRSpec spatial pixel scale of 0.1"/pixel) to maximise the signal-to-noise ratio of the final spectra.

For most uses of the extraction performed on a 3-pixel aperture, the measurements are performed over small wavelength ranges and further corrections for losses due to the smaller extraction box are not required. In the case of full spectral fitting, we use the extraction over the 5- pixel aperture with one exception, JADES-GS-z13-0. In this case we mitigate the wavelength-dependent losses with simultaneous fitting to photometry (see Section 3).

## 2. Empirical measurements

The central aspects of the astrophysical analysis of the spectra has been presented in the main part of the paper. Here we explore a few more issues that bolster the fidelity and robustness of our analysis.

### 2.1 Balmer break index

While the observed spectral breaks in the four objects presented here are fully consistent with expectations for high-redshift galaxies, it is important to test the possibility that the observed breaks may be Balmer breaks at lower redshift. We test this using empirical spectral indices. We adapt the classical Balmer-break index definition[50] and define the break amplitude as the ratio of $f_\lambda$ in the rest-frame range 3751 Å - 4198 Å to that in the range 3145 Å - 3563 Å. This definition expands the spectral windows to include more spectral pixels (15-19 depending on redshift) and increase the signal-to-noise on the index. These measurements are taken from spectra extracted from 3 pixels, maximising the S/N. The effect of wavelength-dependent extraction losses should be minimal for this measurement. When the lower spectral range yields a negative flux, we adopt the 2σ upper flux limit instead. Additionally, to account for the noisy measurement of the (physically) positive-definite flux in the longer wavelength band, we subtract 1σ from the measured flux. This is a conservative upper limit which we quote as a 2σ upper limit in Table 1.

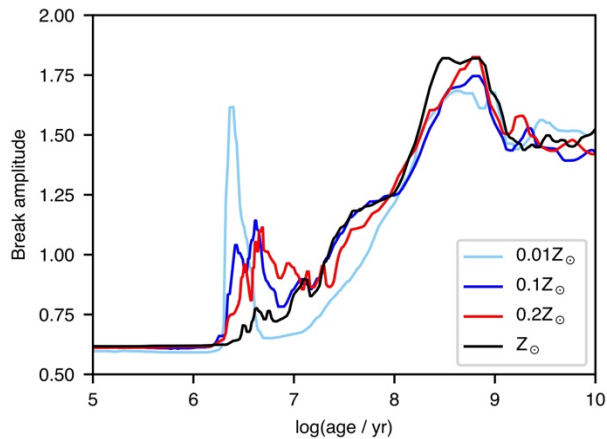

**Extended Data Fig. 1| Balmer-break amplitude plotted against age for single stellar populations with metallicities 0.01$Z_\odot$, 0.1$Z_\odot$, 0.2$Z_\odot$, and $Z_\odot$ (as indicated), according to the models described in Section 3**. The break is defined as the ratio of the flux $f_\lambda$ integrated over the rest-frame 3751–4198Å wavelength range to that in the rest-frame 3145–3563 Å wavelength range. The peak at early ages for all metallicities arises from the onset or red supergiant stars, and that around $6 \times 10^8$ yr from bright asymptotic-giant-branch stars.

In Extended Data Figure 1 we show the evolution of the modified Balmer break index with age for single stellar population models at four different metallicities from 0.01$Z_\odot$ to 1$Z_\odot$. The maximum value reached within 10 Gyr is less than 2.0. We report 2σ lower limits on the value of this index for each of our four targets in Table 1. The smallest measured break strength is measured for GS-z10-0, with a 2σ lower limit of 2.04, which is higher than the maximum reached by the single stellar populations. Again, this suggests that the Balmer break solution is unlikely, although by a smaller margin than for the other three targets.

We note that incorrect background subtraction due to contamination of background shutters by neighbouring galaxies or the source itself can lead to a biased measurement of the break strength. We are careful with the reduction to exclude any contaminated shutters in the background estimate, as described in Methods section 1, and so do not expect these measurements to be affected in this way.

## 2.2 Limits on possible emission lines

Apart from the possible detections of Lyα in JADES-GS-z10-0 and [Ne III]λ3869 in JADES-GS-z12-0 (which we will assess in future work), visual inspection of the 1D and 2D spectra did not show the presence of any emission lines above the level of the noise in any of the four targets. We derive upper limits to emission-line fluxes and equivalent widths using the error spectrum output from the data reduction pipeline for the optimised S/N spectra extracted over 3 pixels. Our reduced spectra have an irregular wavelength grid, and we estimate that the line spread function of these PRISM spectra results in unresolved emission lines with FWHM of approximately 2-3 spectral pixels. Thus, to calculate emission-line limits, for the 3 spectral pixels centred on the expected centroid of the emission line at the calculated redshift, we sum the pixel errors in quadrature and multiply the result by the wavelength interval between pixels. This results in 1σ upper limits on line fluxes, which we convert into equivalent-width limits by fitting a simple polynomial to the continuum of each object to get the level on the continuum and associated uncertainty. Table 1 reports 2σ upper limits on the equivalent widths of He IIλ1640, C III]λλ1907,1909 and [O II]λλ3726,3729 while the full set of limits on rest-frame UV emission lines is given in Extended Data Table 1.

| JADES-ID: | GS-z10-0 | GS-z11-0 | GS-z12-0 | GS-z13-0 |
| --- | --- | --- | --- | --- |
| EW(C III]λλ1907, 1909) | < 13.74 | < 5.84 | < 11.94 | < 14.40 |
| EW(He IIλ1640) | < 14.80 | < 5.98 | < 12.84 | < 14.52 |
| EW(C IVλ1548) | < 13.92 | < 6.08 | < 12.98 | < 14.20 |
| EW(O III]λλ1661,1666) | < 14.12 | < 5.94 | < 12.80 | < 14.74 |
| EW([O II]λλ3726,3729) | < 28.10 | < 9.07 | < 16.62 | < 16.82 |
| C III]λλ1907,1909 | < 8.43e-20 | < 5.60e-20 | < 4.88e-20 | < 7.09e-20 |
| He IIλ1640 | < 1.26e-19 | < 8.08e-20 | < 7.14e-20 | < 1.01e-19 |
| C IVλλ1548,1550 | < 1.33e-19 | < 9.23e-20 | < 7.98e-20 | < 1.11e-19 |
| O III]λλ1661,1666 | < 1.17e-19 | < 7.77e-20 | < 6.92e-20 | < 9.93e-20 |
| [O II] λλ3726,3729 | < 2.99e-20 | < 2.86e-20 | < 5.09e-20 | < 5.29e-20 |

**Extended Data Table 1** | 2σ upper limits on the rest-frame equivalent widths (in Å) and observed line fluxes (in erg $s^{-1}$ $cm^{-2}$) of rest-frame UV emission lines.

## 2.3 $M_{UV}$ and UV slopes

The UV slope, β, was determined directly from the 1D extracted spectra. We performed a least-squares fit to the gradient in the ln(λ):ln($f_\lambda$) space, with the errors on ln($f_\lambda$) taken to be ln($f_\lambda + \sigma$) − ln($f_\lambda$) for each extracted spectral pixel (where σ is the noise). For all objects we fit the β slope over the rest-frame wavelength range 1250 – 2600 Å, following ref. 51 (fitting to the entire wavelength range, since absorption features avoided with the spectral windows of ref. 51 will not significantly affect the measurement at this level of S/N and resolution), with the exception of JADES-GS-z10-0 where we used a slightly smaller range of 1500 – 2600 Å to avoid the possible Lyα emission and damping wing. These results were also consistent with those from fitting a power-law to $f_\lambda \propto \lambda^{-\beta}$ in linear wavelength space, weighting each point by $1/\sigma^2$.

We determined the absolute magnitude in the rest-frame UV ($M_{UV}$) around $\lambda_{rest}$ = 1500Å by measuring the average flux density per unit frequency interval ($f_\nu$) from the extracted spectra over the rest-frame wavelength range 1400 – 1600 Å, and accounting for luminosity distance. The errors on each individual pixel flux density were combined in quadrature to derive the uncertainty in β.

We note that $M_{UV}$ measured from the spectrum alone is somewhat fainter and the β somewhat redder for JADES-GS-z12-0 than that measured from NIRCam data in the companion paper[27]. We attribute this to residual slit losses missed by our standard correction, since this object is quite close to the edge of the shutter. This highlights the importance of the complementarity between NIRSpec spectroscopy and NIRCam photometry. The difference is starker for JADES-GS-z13-0, which is very close to the edge of the shutter, and for which we report instead $M_{UV}$ and β derived from SED modelling including the NIRCam photometry (see Section 3). The comparison with the measurements in ref. 27 shows good agreement for JADES-GS-z10-0 and JADES-GS-z11-0, which validates the spectroscopic flux calibration in the case where the adoption of point-source path losses is a good approximation.

## 3. BEAGLE SED fitting

We perform full spectral fitting to the R100 spectra using the BEAGLE code[37]. In general, firm constraints on metallicities (both nebular and stellar) and gas parameters require the spectroscopic detection of emission and absorption lines, while constraints on total stellar masses and star-formation rates generally require NIRCam photometry, since NIRSpec MSA spectra are so prone to slit losses. As these objects are so small [$r_{1/2} \simeq 50 - 165$ pc, with on-sky sizes of $\theta_{1/2} \simeq 0.015–0.04"$ [27]] we use pre-calculated point-source slit-loss corrections. Therefore, in this paper we present an entirely independent analysis to that in ref. 27, both in datasets used (except for the fits to JADES-GS-z13-0 where we employ the NIRCam photometry, see later in this section for details), in SED codes and parameterizations. We comment on the consistency between the two analyses throughout this section.

This requires modelling of the wavelength-dependent line-spread function (LSF). We fit Gaussian profiles to emission lines in R100 spectra taken within this deep pointing and compare their widths as a function of wavelength to the dispersion curves provided by STScI (https://jwst-docs.stsci.edu/jwst-near-infrared-spectrograph/nirspec-instrumentation/ nirspec-dispersers-and-filters). We find that the supplied dispersion curves multiplied by a factor of 0.7 provide a reasonable representation of the measured wavelength-dependent LSF.

| Parameter | Description | Prior |
|---|---|---|
| $\log(M_{tot}/M_\odot)$ | Logarithm of the integrated SFH. | Uniform $\in [6, 12]$ |
| $M/M_\odot$ | Stellar mass, including stellar remnants. | Not fitted (dependent on $M_{tot}$, Z and t) |
| $\psi/M_\odot yr^{-1}$ | Current star formation rate. | Not fitted (dependent on $M_{tot}$ and t) |
| $\log(Z/Z_\odot)$ | Logarithm of metallicity of stars. | Uniform $\in [-2.2, 0.4]$ |
| $\log(Z_{gas}^{HII}/Z_\odot)$ | Metallicity of gas in H ii regions. | Set equal to Z |
| $\hat{\tau}_v$ | Total V-band attenuation optical depth in the interstellar medium (ISM). | Exponential $\in [0, 6]$ |
| μ | Fraction of $\hat{\tau}_v$ arising from dust in the diffuse ISM | Fixed to 0.4. |
| $\log U_s$ | Effective gas ionization parameter in H ii regions | Dependent on $Z_{gas}^{HII}$ (following Eq. 1) |
| $\xi_d$ | Dust-to-metal mass ratio in H ii regions. | Fixed to 0.1 |
| $n_H/cm^{-3}$ | Hydrogen gas density in H ii regions. | Fixed to 100. |
| (C/O)/(C/O)$_\odot$ | Carbon-to-oxygen abundance ratio in units of (C/O)$_\odot$ = 0.44 | Fixed to unity. |
| $m_{up}/M_\odot$ | Upper mass cutoff of the IMF | Fixed to 100. |
| Log(t/Gyr) | Logarithm of the age of the oldest stars. | Uniform $\in [6, 10.8]$ |
| z | redshift | Gaussian $\mu_z = z_{1216}$, $\sigma_z = 0.01$ |

**Extended Data Table 2 |** Parameter descriptions and prior distributions used in BEAGLE fitting.

We use spectra extracted over the full micro-shutter aperture to minimise the effects of wavelength dependent losses, since the size of the shutter is more than twice the width of the PSF at 5μm. JADES-GS-z13-0, however, is just at the edge of the shutter and the 2D spectrum shows that it is clearly truncated. Therefore, to provide information of the aperture losses for this target, we use a 3-pixel extraction box to maximise S/N and simultaneously fit to the NIRCam aperture photometry[27]. We multiply the shape of the spectrum by a second-order polynomial, sampling over the polynomial coefficients in the fit. This essentially allows the NIRCam photometry to set the normalisation of the spectrum while also correcting wavelength-dependent slit-losses in the spectral calibration. The fits and associated parameter constraints are shown in Extended Data Figures 2-3. We note that SED fitting is performed in the companion paper[27], yet there they fit only to NIRCam photometry, fixing the redshift to the spectroscopic redshift. Here we perform a complimentary analysis, fitting only to the spectra (except for JADES-GS-z13-0, as explained above).

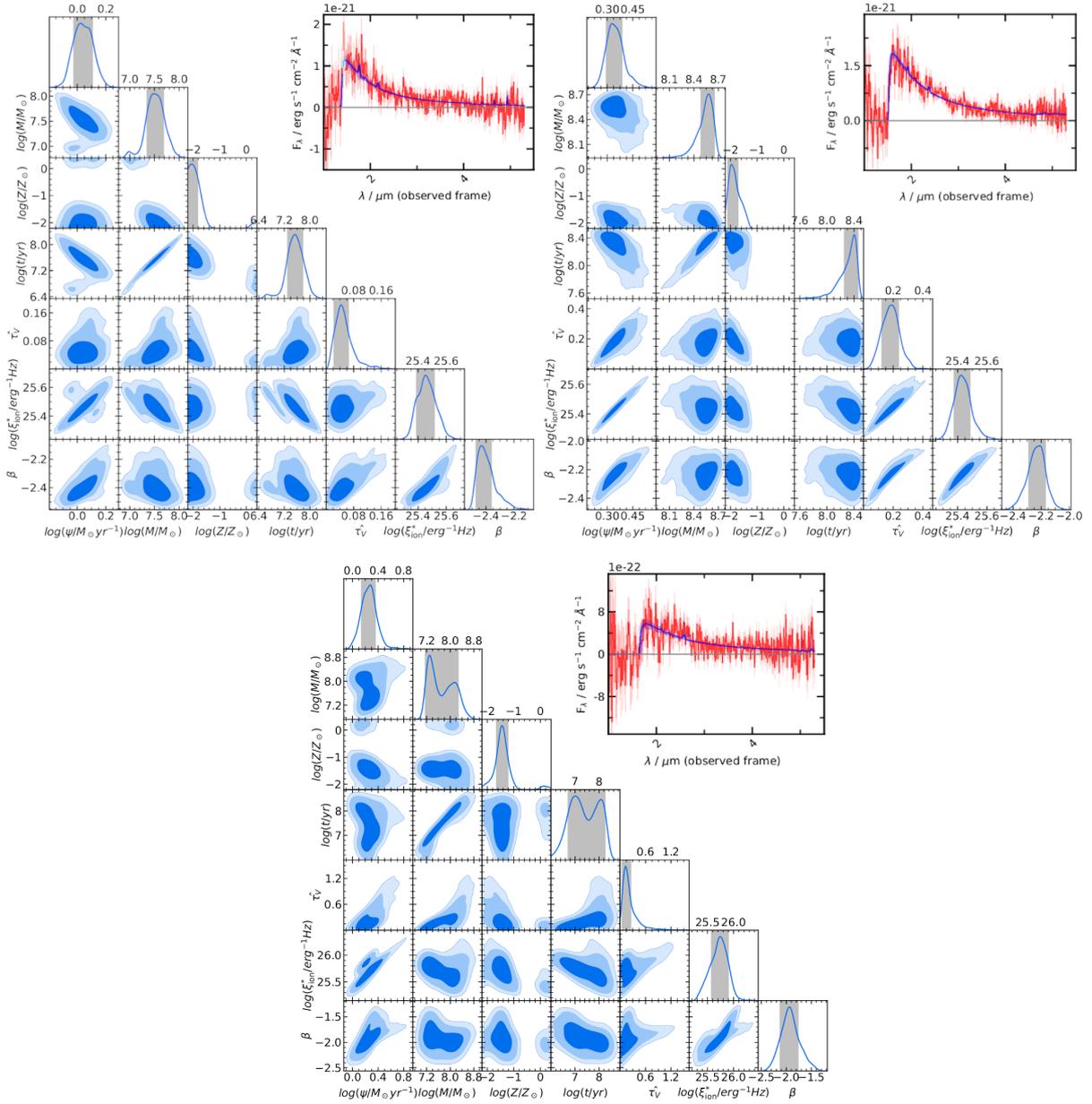

**Extended Data Fig. 2|** The results of full spectral fitting to JADES-GS-z10-0 (top left), JADES-GS-z11-0 (top right) and JADES-GS-z12-0 (bottom) with BEAGLE. We fit models to spectra extracted over the full shutter aperture to minimise the wavelength-dependent losses due to varying point-spread function (PSF). The triangle plot shows the 2D (off-diagonal) and 1D (along the main diagonal) posterior probability distributions on stellar mass (M), metallicity (Z), maximum age of stars (t) and the effective dust attenuation optical depth in the V-band ($\hat{\tau}_V$) which are all derived from the beagle fits. We also include the model constraints on the star-formation rate ($\Psi$), UV slope ($\beta$) and ionising photon emissivity ($\xi_{ion}$), which are derived parameters of the model. The dark, medium and light blue contours show the extents of the 1, 2 and 3σ credible regions of the posterior probability, respectively. The inset panel shows the observed spectrum and 1σ standard errors per pixel in red and light red respectively, and the median and 1σ range in fitted model spectra in blue. We fit with a constant star formation history (more details in the text and Methods section 3).

For our BEAGLE fits, we mask the region of possible Lyα in JADES-GS-z10-0 (between 1.4148 and 1.4509μm, inclusive), since it is offset from the break, and would require specialised modelling of the line shape and offset to be accounted for properly. The masked region is shown as pale blue in the spectrum in Extended Data Figure 2. (upper right) For JADES-GS-z12-0, we mask regions of rest-frame UV emission lines (light blue regions in Extended Data Figure 2, lower panel, covering 2.1081 to 2.1620μm, 2.2875 to 2.3181μm and 2.6261 to 2.6533μm) since noise structure in the spectrum is over-fitted if left un-masked.

We then fit the spectra following the procedure of ref. 52. We use a constant SFH and fixed nebular parameters since we see no emission lines. The list of parameters employed in the fits, as well as chosen priors are given in Extended Data Table 2. We use the updated Bruzual & Charlot stellar population synthesis templates[53], as described in ref. 54 with the physically consistent nebular line+continuum emission grid of ref. 55. We adopt a Chabrier[56] initial mass function with an upper mass limit of 100M$_\odot$.

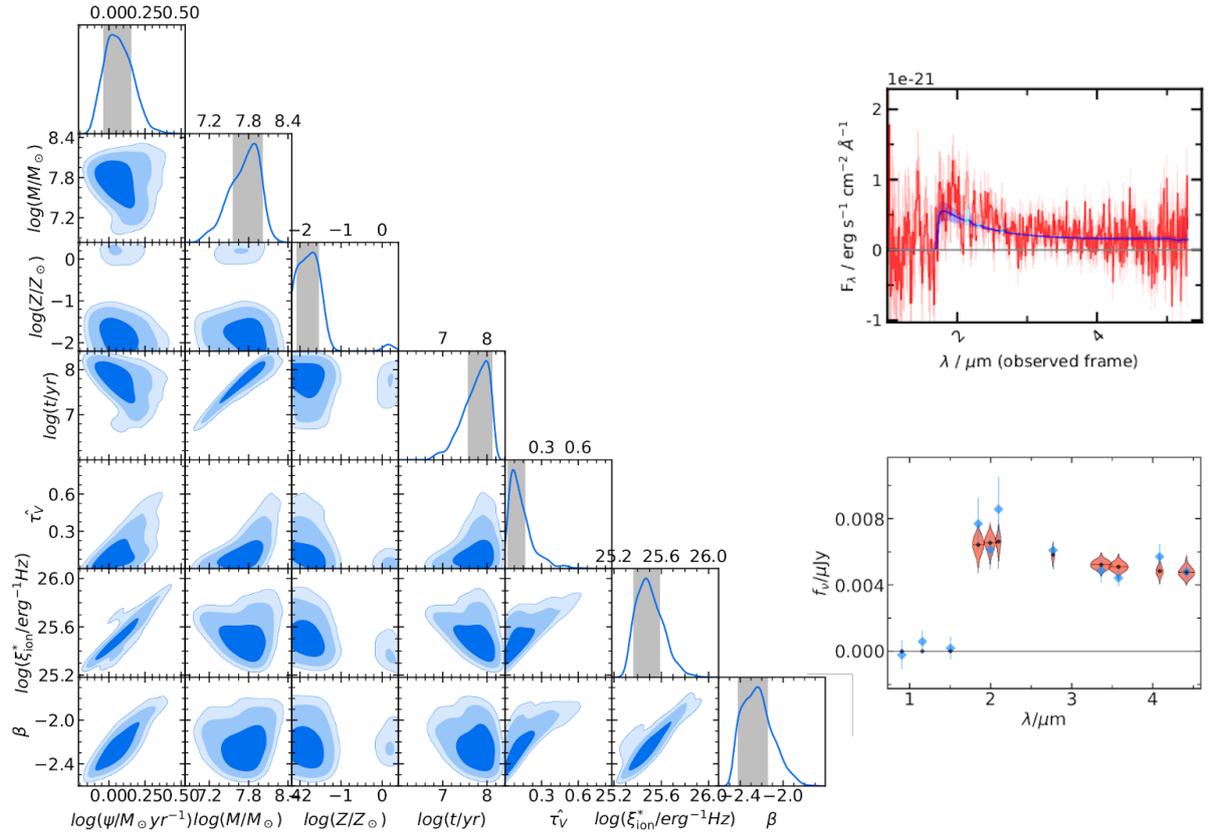

**Extended Data Fig. 3** | As for Extended Data Figure 2, but for BEAGLE fits to JADES-GS-z13-0. The bottom right panel shows the observed photometry and associated as blue diamonds and associated 1σ s.d. error-bars while the coral shaded regions show the model photometry in the same bands. Since this galaxy is very close to the edge of the shutter, we use an extraction over 3 pixels to maximise the S/N. Then to account for wavelength-dependent slit losses we simultaneously fit the spectrum and NIRCam photometry.

We have verified that the results are not significantly changed when assuming an upper mass limit of 300M$_\odot$. We account for the depletion of metals onto dust grains in the photoionized interiors of stellar birth clouds and include attenuation by dust in the outer neutral envelopes of the clouds and in the diffuse ISM[57]. We set the ionisation parameter, log U$_s$, to depend on the nebular metallicity (and hence stellar metallicity) according to:

$$log\ U_s = -3.638 + 0.055Z + 0.68Z^2$$

which follows the observations of ref. 58. The results do not change significantly for JADES-GS-z10-0 and JADES-GS-z11-0 when logU$_s$ is allowed to vary freely in the range −4 < log U$_s$ < −1. However, Z and log U$_s$ are unconstrained in JADES-GS-z13-0, and poorly constrained in JADES-GS-z12-0, when log U$_s$ is allowed to vary freely.

To test what is driving the fits to very low metallicity, we fit fixing the metallicity to intermediate values (between Solar and 10% of solar). In this range large [O II]λλ3726,3729 model fluxes are not described well by the data, and the derived constraints are pushed to low or high metallicities in the two lowest redshift sources. We note, however, that reasonable intermediate metallicity solutions can be fit to the two higher redshift galaxies. Moreover, letting log $U_s$ vary freely in this metallicity range decreases model [O II]λλ3726,3729 fluxes but increases the model C III]λλ1907,1909 fluxes, constraining the fits still to the edges of the prior for the two lowest redshift galaxies. It is a complex interplay in the limiting fluxes of these emission lines that drive the low metallicity constraints in these galaxies.

Since the constraints on the stellar metallicity are driven by the lack of strong emission lines, we tested whether recent cessation of star formation would significantly change the constraints. We therefore tested a constant star formation history where the SFR of the recent 10 Myr was allowed to vary independently (and decrease). We find that the star-formation rate, Ψ, is fairly unconstrained with low posterior median values, meaning recent cessation is consistent with the data. However, we still infer very low metallicities in JADES-GS-z10-0 and JADES-GS-z11-0 (the two with highest S/N spectra). For the two lower S/N spectra (JADES-GS-z12-0 and JADES-GS-z13-0), Z, $\hat{\tau}_v$ and Ψ are very poorly constrained when this extra free parameter is included. We show the results for JADES-GS-z10-0 and JADES-GS-z11-0 in Extended Data Table 3.

|  | *JADES-GS-z10-0* | | *JADES-GS-z11-0* | |
| --- | --- | --- | --- | --- |
|  | $SFR_{10}$ varied | $f_{esc}$ varied | $SFR_{10}$ varied | $f_{esc}$ varied |
| $log(M/M_\odot)$ | $7.8^{+0.12}_{-0.11}$ | $7.41^{+0.1}_{-0.11}$ | $8.85^{+0.12}_{-0.15}$ | $8.70^{+0.11}_{-0.16}$ |
| $\psi/M_\odot yr^{-1}$ | $-2.11^{+1.28}_{-1.34}$ | $2.43^{+0.65}_{-0.58}$ | $-1.87^{+1.45}_{-1.37}$ | $2.83^{+0.76}_{-0.47}$ |
| $log(t/yr)$ | $7.15^{+0.17}_{-0.11}$ | $6.29^{+0.19}_{-0.2}$ | $8.12^{+0.20}_{-0.28}$ | $8.26^{+0.14}_{-0.23}$ |
| $log(Z/Z_\odot)$ | $-1.88^{+0.36}_{-0.21}$ | $-1.7^{+0.41}_{-0.33}$ | $-1.87^{+0.33}_{-0.21}$ | $-1.8^{+0.29}_{-0.25}$ |
| $\hat{\tau}_v$ | $0.3^{+0.26}_{-0.18}$ | $0.09^{+0.05}_{-0.04}$ | $0.68^{+0.18}_{-0.2}$ | $0.26^{+0.1}_{-0.08}$ |
| $\xi_{ion}$ | $25.04^{+0.17}_{-0.21}$ | $26.01^{+0.12}_{-0.13}$ | $24.73^{+0.25}_{-0.15}$ | $25.54^{+0.12}_{-0.08}$ |
| $f_{esc}$ | — | $0.86^{+0.1}_{-0.13}$ | — | $0.64^{+0.24}_{-0.32}$ |

**Extended Data Table 3 |** BEAGLE-derived parameters when additionally fitting the star formation rate in the last 10 Myr allowed to vary freely independently of the previous history (labelled SFR$_{10}$ varied), or when varying the escape fraction of Lyman-continuum $f_{esc}$ (labelled $f_{esc}$ varied). We show the results when fitting to the two objects with highest S/N, JADES-GS-z10-0 and JADES-GS-z11-0 because the constraints on the other two objects are very poor when adding an extra free parameter to the fits.

Another possibility to explain relatively weak line emission is a high escape of Lyman-continuum photons from the galaxy. We fit with a picket fence model [allowing for clear sight-lines to the stars through the outer neutral envelopes of birth clouds[59]]. The results are also given in Extended Data Table 3 for the two objects (JADES-GS-z10-0 and JADES-GS-z11-0) with the highest S/N spectra. We note that JADES-GS-z10-0 shows a solution with very low age (a few Myr) and high escape fraction. The measured metallicity is marginally higher in this case, but still very low within the 1σ credible interval. JADES-GS-z11-0 does not show such a peak in the posterior distribution function, with fits still favouring older ages and escape fractions that span the input uniform prior. We note that these results are consistent with ref. 27, who find similarly low age and high escape fraction for JADES-GS-z10-0 compared to JADES-GS-z11-0.

We also fit the spectra assuming the main feature is a Balmer break (see also Sec. 2). The results are shown in Extended Data Figure 4. Here, we fit a delayed SFH which halted a Gyr prior to observation, varying metallicity, maximum stellar age and redshift within a tight prior centred on the assumed redshift in the case that the break is a Balmer break. We see that the fits consistently fail to reproduce the peak and blue slope red-ward of the break, showing poor spectral fits. Additionally, JADES-GS-z10-0 and JADES-GS-z11-0 show flux blue-ward of the break in the fitted models which is clearly inconsistent with the measured flux and noise limits.

We further explored possible low redshift solutions to the photometry by SED-fitting with BEAGLE-AGN[60], which includes narrow-line emission from obscured active-galactic-nuclei. This type of template is rarely used when fitting to high-redshift galaxy candidates. Three of the four galaxies provide intermediate redshift (z∼3–3.5) solutions to the photometry, and the fit to JADES-GS-z13-0 is displayed in Extended Data Fig. 5. We see no strong emission lines in the spectra themselves, hence disproving these low-redshift possibilities allowed by the photometry.

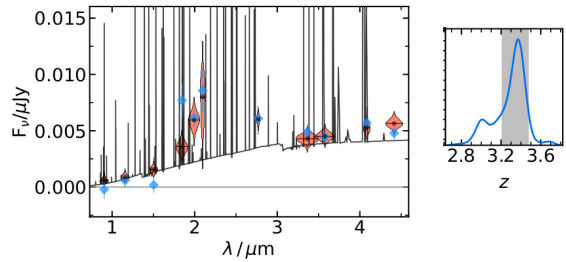

**Extended Data Fig. 5 |** The left panel shows the measured NIRCam photometry and associated *1σ* s.d. error bars for JADES-GS-z13-0 as blue diamonds and lines, respectively. The coral violin shaded regions show the underlying model values. The black line shows the maximum a-posteriori probability solution with strong emission lines due to active-galactic-nuclei narrow-line emission. The right panel shows a zoom of the intermediate redshift solution fitted to the photometry.

## 4. Damping wing profile

Galaxies at the very large redshifts presented in this paper are embedded in a largely neutral IGM, which has not yet undergone reionization. In this case the effective optical depth of the hydrogen at lower redshifts along the line of sight becomes so large that the accumulated absorption in the Lorentzian scattering wing of the Lyman alpha resonance line causes intergalactic absorption to spill over into wavelengths above the rest frame Lyα line. This so-called damping wing absorption softens the sharp cutoff in the spectrum due to the intervening intergalactic hydrogen.

We have included the effect of the damping wing absorption in our spectral fits using the prescription presented by Miralda-Escudé[61], who first pointed out the important effect. The model assumes that the damping wing arises in a uniformly distributed completely neutral IGM containing the bulk of the baryons in the universe. For a source at a given redshift $z_s$ it has only two free parameters, $\tau_0$, the overall strength of the Lyman alpha absorption in the form of the optical depth of the classical Gunn-Peterson trough[62] at a reference redshift of z = 5, and $z_n$, the redshift below which the intergalactic hydrogen is assumed to abruptly transition from fully neutral to fully ionized.

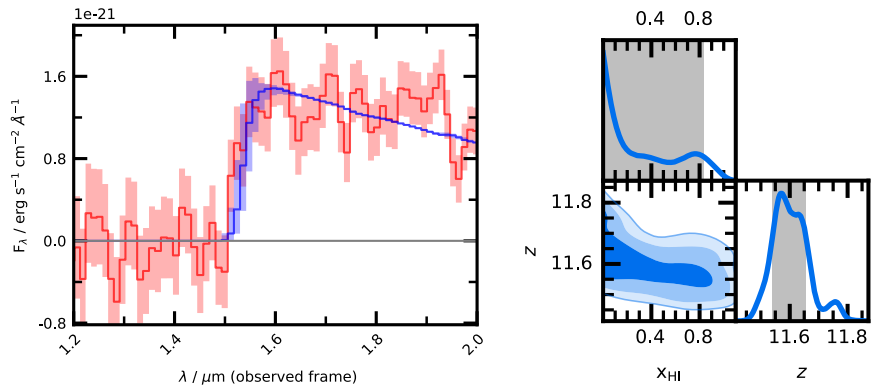

**Extended Data Fig. 6 |** As for Fig. 3 but showing a fit to the damping wing using a different definition of 'best fit' to fix the physical parameters of the galaxy spectrum which pushes the constraints to a higher-redshift solution.

These two parameters are in turn set by the assumed cosmological model, which we here take to be Planck 2015 ΛCDM[63]. This model's baryonic density parameter of $\Omega_b h$ = 0.033, total mass density parameter $\Omega_M$ = 0.309, together with the primordial Helium abundance of Y = 24% translate into $\tau_0$ = $3.1 \times 10^5$. The Planck satellite also measured $z_n$ = 8.8 for this cosmology, although the predicted damping wing absorption is insensitive to this parameter at the large z > 10 redshifts relevant here. A partially reionized IGM is included in the customary manner by multiplying $\tau_0$ by the volume-averaged neutral fraction $x_{hi}$. This simple model ignores the potential complications of the galaxies being observed displaying strong intrinsic neutral hydrogen absorption, or their having already reionized a large volume of their immediate surroundings[64].

In fitting the damping wing to JADES-GS-z11-0, we found a bimodality in the redshift solution, with a higher redshift solution being consistent with a fully ionised IGM. We find that the final solution is quite sensitive to the adopted intrinsic underlying spectrum. We find solutions at higher redshifts provide poorer fits to the break region itself (shown in Extended Data Fig 6 for completeness). The difference between fits was based on the definition of 'best fit' to the full spectrum used to fix physical parameters (either the minimum chi-2 solution, Fig. 3, or the maximum a-posteriori probability solution shown here), while if we do not model Lyman-alpha emission in the fitting, we find a bimodal solution with redshift.

**Data availability** The data that support the findings of this study are available from the corresponding author upon reasonable request.

**Code availability** BEAGLE is available via a Docker image (distributed through docker hub) upon request at https:/iap.fr/beagle.

**Acknowledgments.** For the purpose of open access, the author has applied a Creative Commons Attribution (CC BY) licence* to any Author Accepted Manuscript version arising. ECL acknowledges support of an STFC Webb Fellowship (ST/W001438/1). SC acknowledges support by European Union's HE ERC Starting Grant No. 101040227 - WINGS. MC, FDE, TJL, RM, JW, and LS acknowledge support by the Science and Technology Facilities Council (STFC), ERC Advanced Grant 695671 "QUENCH". RM is further supported by a research professorship from the Royal Society. JW is further supported by the Fondation MERAC. HÜ gratefully acknowledges support by the Isaac Newton Trust and by the Kavli Foundation through a Newton-Kavli Junior Fellowship. NB and PJ acknowledge support from the Cosmic Dawn Center (DAWN), funded by the Danish National Research Foundation under grant no.140. RS acknowledges support from a STFC Ernest Rutherford Fellowship (ST/S004831/1). AJB, AJC, JC, IEBW, AS, & GCJ acknowledge funding from the "FirstGalaxies" Advanced Grant from the European Research Council (ERC) under the European Union's Horizon 2020 research and innovation pro- gramme (Grant agreement No. 789056). BER, BDJ, DJE, MR, EE, GR, CNAW, and FS acknowledge support from the JWST/NIRCam Science Team contract to the University of Arizona, NAS5-02015. DJE is further supported as a Simons Investi- gator. RB acknowledges support from an STFC Ernest Rutherford Fellowship [grant number ST/T003596/1]. REH acknowledges support from the National Science Foundation Graduate Research Fellowship Program under Grant No. DGE-1746060. SAr, BRP, and MP acknowledge support from the research project PID2021-127718NB- I00 of the Spanish Ministry of Science and Innovation/State Agency of Research (MICIN/AEI). MP is further supported by the Programa Atracción de Talento de la Comunidad de Madrid via grant 2018-T2/TIC-11715. LW acknowledges support from the National Science Foundation Graduate Research Fellowship under Grant No. DGE-2137419. KB is supported in part by the Australian Research Council Cen- tre of Excellence for All Sky Astrophysics in 3 Dimensions (ASTRO 3D), through project number CE170100013. RH was funded by the Johns Hopkins University, Insti- tute for Data Intensive Engineering and Science (IDIES). This research made use of the lux supercomputer at UC Santa Cruz, funded by NSF MRI grant AST 1828315. Acknowledgement for getting assigned a protected node for the DEEP BagPipes runs: "This study made use of the Prospero high performance computing facility at Liverpool John Moores University."

**Author contributions.** ECL and SCa led the writing of this paper. MR, CNAW, EE, FS, KH, CCW contributed to the design, construction, and com- missioning of NIRCam. AB, AD, CNAW, CW, DJE, H-WR, MR, MF, PF, PJ, RM, SAl, SAr contributed to the design of the JADES survey. BER, ST, BDJ, CNAW, DJE, IS, MR, RE, ZC contributed to the JADES imaging data reduction. RHa, BER contributed to the JADES imaging data visualization. BDJ, ST, AD, DPS, LW, MWT, RE contributed the modeling of galaxy photometry. KH, JMH, JL, LW, RE, REH contributed the photometric redshift determination and target selection. BDJ, EN, KAS, ZC contributed to the JADES imaging morphological analysis. BER, CNAW, CCW, KH, MR contributed to the JADES pre-flight imaging data challenges. SCa, MC, JW, PF, GG, SAr, BRdP, contributed to the NIRSpec data reduction and to the development of the NIRSpec pipeline PJ, NB, SAr contributed to the design and optimisation of the MSA configurations. AJC, AB, CNAW, ECL, HU, RB, KB, contributed to the selection, prioritisation and visual inspection of the targets. SCh, JC, ECL, RM, JW, RS, FDE, MM, MC, AdG, GJ, AS, LS contributed to analysis of the spectroscopic data, including redshift determination and spectral modelling. PJ, PF, MS, TR, GG, NL, NK, BRdP contributed to the design, construction and commissioning of NIRSpec. FDE, TL, MM, MC, BRdP, RM, SAr contributed to the development of the tools for the spectroscopic data analysis, visualisation and fitting. CW contributed to the design of the spec- troscopic observations and MSA configurations. BER, CW, DJE, DPS, MR, NL, and RM serve as the JADES Steering Committee.

**Competing Interests Statement.** The authors declare no competing interests

# References

1. Dayal, P., Ferrara, A.: Early galaxy formation and its large-scale effects. Physics Reports 780, 1–64 (2018) https://arxiv.org/abs/1809.09136 [astro-ph.GA]. https://doi.org/10.1016/j.physrep.2018.10.002
2. Bromm, V., Coppi, P.S., Larson, R.B.: The Formation of the First Stars. I. The Primordial Star-forming Cloud. Astrophys. J. 564(1), 23–51 (2002) https://arxiv.org/abs/astro-ph/0102503 [astro-ph]. https://doi.org/10.1086/323947
3. Schneider, R., Ferrara, A., Salvaterra, R.: Dust formation in very massive primordial supernovae. Mon. Not. R. Astron. Soc. 351(4), 1379–1386 (2004) https://arxiv.org/abs/astro-ph/0307087 [astro-ph]. https://doi.org/10.1111/j.1365-2966.2004.07876.x
4. Jeon, M., Bromm, V., Pawlik, A.H., Milosavljević, M.: The first galaxies: simulating their feedback-regulated assembly. Mon. Not. R. Astron. Soc. 452(2), 1152–1170 (2015) https://arxiv.org/abs/1501.01002 [astro-ph.GA]. https://doi.org/10.1093/mnras/stv1353
5. Vogelsberger, M., et al.: High-redshift JWST predictions from IllustrisTNG: dust modelling and galaxy luminosity functions. Mon. Not. R. Astron. Soc. 492(4), 5167–5201 (2020) https://arxiv.org/abs/1904.07238 [astro-ph.GA]. https://doi.org/10.1093/mnras/staa137
6. Hutter, A., et al.: Astraeus I: the interplay between galaxy formation and reionization. Mon. Not. R. Astron. Soc. 503(3), 3698–3723 (2021) https://arxiv.org/abs/2004.08401 [astro-ph.GA]. https://doi.org/10.1093/mnras/stab602
7. Wilkins, S.M., et al.: First Light And Reionisation Epoch Simulations (FLARES) V: The redshift frontier. Mon. Not. R. Astron. Soc. 519(2) 3118-3128 (2023) https://arxiv.org/abs/2204.09431 [astro-ph.GA] https://doi.org/10.1093/mnras/stac3280
8. Steidel, C.C., Giavalisco, M., Pettini, M., Dickinson, M., Adelberger, K.L.: Spectroscopic Confirmation of a Population of Normal Star-forming Galaxies at Redshifts $Z > 3$. Astrophys. J. Lett. 462, 17 (1996) https://arxiv.org/abs/astro-ph/9602024 [astro-ph]. https://doi.org/10.1086/310029
9. Madau, P., et al.: High-redshift galaxies in the Hubble Deep Field: colour selection and star formation history to $z\sim4$. Mon. Not. R. Astron. Soc. 283(4), 1388–1404 (1996) https://arxiv.org/abs/astro-ph/9607172 [astro-ph]. https://doi.org/10.1093/mnras/283.4.1388
10. Steidel, C.C., et al.: Lyman Break Galaxies at Redshift $z \sim 3$: Survey Description and Full Data Set. Astrophys. J. 592(2), 728–754 (2003) https://arxiv.org/abs/astro-ph/0305378 [astro-ph]. https://doi.org/10.1086/375772
11. Zavala, J.A., et al.: A dusty starburst masquerading as an ultra-high redshift galaxy in JWST CEERS observations. Astrophys. J. L. 943(2), L9 (2023) https://arxiv.org/abs/2208.01816 [astro-ph.GA]
12. Williams, H. et al: Spectroscopy from Lyman alpha to [O III] 5007 of a Triply Imaged Magnified Galaxy at Redshift z = 9.5 arXiv e-prints, 2210–15699 (2022) https://arxiv.org/abs/2210.15699 [astro-ph.GA]
13. Oesch, P.A., et al.: A Remarkably Luminous Galaxy at z=11.1 Measured with Hubble Space Telescope Grism Spectroscopy. Astrophys. J. 819(2), 129 (2016) https://arxiv.org/abs/1603.00461 [astro-ph.GA]. https://doi.org/10.3847/0004-637X/819/2/129
14. Jiang, L., et al.: Evidence for GN-z11 as a luminous galaxy at redshift 10.957. Nature Astronomy 5, 256–261 (2021) https://arxiv.org/abs/2012.06936 [astro-ph.HE]. https://www.nature.com/articles/s41550-020-01275-y
15. Donnan, C.T., et al.: The evolution of the galaxy UV luminosity function at redshifts $z \sim 8$-15 from deep JWST and ground-based near-infrared imaging. Mon. Not. R. Astron. Soc. 518(4), 6011-6040, (2023) https://arxiv.org/abs/2207.12356 [astro-ph.GA]
16. Harikane, Y., et al.: A Search for H-Dropout Lyman Break Galaxies at z 12-16. Astrophys. J. 929(1), 1 (2022) https://arxiv.org/abs/2112.09141 [astro-ph.GA]. https://doi.org/10.3847/1538-4357/ac53a9
17. Adams, N.J., et al.: Discovery and properties of ultra-high redshift galaxies (9 < z < 12) in the JWST ERO SMACS 0723 Field. Mon. Not. R. Astron. Soc. 518(3), 4755-4766 (2023) https://arxiv.org/abs/2207.11217 [astro-ph.GA]
18. Finkelstein, S.L., et al.: CEERS Key Paper I: An Early Look into the First 500 Myr of Galaxy Formation with JWST. arXiv e-prints, 2211–05792 (2022) https://arxiv.org/abs/2211.05792 [astro-ph.GA]
19. Finkelstein, S.L., et al.: A Long Time Ago in a Galaxy Far, Far Away: A Candidate $z \sim 12$ Galaxy in Early JWST CEERS Imaging. Astrophys. J. Lett. 940(2), L55 (2022) https://arxiv.org/abs/2207.12474 [astro-ph.GA]

20. Castellano, M., et al.: Early Results from GLASS-JWST. III. Galaxy Candidates at z 9-15. Astrophys. J. Lett. 938(2), 15 (2022) https://arxiv.org/abs/2207.09436 [astro-ph.GA]. https://doi.org/10.3847/2041-8213/ac94d0
21. Whitler, L., et al.: On the ages of bright galaxies ~ 500 Myr after the Big Bang: insights into star formation activity at zrsim15 with JWST. Mon. Not. R. Astron. Soc. 519(1), 157-171 (2023) https://arxiv.org/abs/2208.01599 [astro-ph.GA]
22. Atek, H., et al.: Revealing Galaxy Candidates out to z ~ 16 with JWST Observations of the Lensing Cluster SMACS0723. Mon. Not. R. Astron. Soc. 519(1), 1201-1220 (2023) https://arxiv.org/abs/2207.12338 [astro-ph.GA]
23. Bouwens, R.J., et al.: Evolution of the UV LF from z~15 to z~8 Using New JWST NIRCam Medium-Band Observations over the HUD-F/XDF. arXiv e-prints, 2211–02607 (2022) https://arxiv.org/abs/2211.02607 [astro-ph.GA]
24. Bouwens, R.J., et al.: UV Luminosity Density Results at z>8 from the First JWST/NIRCam Fields: Limitations of Early Data Sets and the Need for Spectroscopy. arXiv e-prints, 2212-06683 (2022) https://arxiv.org/abs/2212.06683 [astro-ph.GA]
25. Jakobsen, P., et al.: The Near-Infrared Spectrograph (NIRSpec) on the James Webb Space Telescope. I. Overview of the instrument and its capabilities. Astron. Astrophys. 661, 80 (2022) https://arxiv.org/abs/2202.03305 [astro-ph.IM]. https://doi.org/10.1051/0004-6361/202142663
26. Rieke, M.J., Kelly, D., Horner, S.: Overview of James Webb Space Telescope and NIRCam's Role. Proc. SPIE 5904, 1–8 (2005). https://doi.org/10.1117/12.615554
27. Robertson, B.E., et al.: Discovery and properties of the earliest galaxies with confirmed distances. submitted to Nature Astronomy (2023)
28. Furtak, L.J. et al.: JWST UNCOVER: A triply imaged faint quasar candidate at zphot≃7.7 arXiv e-prints, 2212-10531 (2022) https://arxiv.org/abs/2212.10531 [astro-ph.GA]
29. Bouwens, R.J., et al.: A candidate redshift z~10 galaxy and rapid changes in that population at an age of 500Myr. Nature 469(7331), 504–507 (2011) https://arxiv.org/abs/0912.4263 [astro-ph.CO]. https://doi.org/10.1038/nature09717
30. Ellis, R.S., et al.: The Abundance of Star-forming Galaxies in the Redshift Range 8.5-12: New Results from the 2012 Hubble Ultra Deep Field Campaign. Astrophys. J. Lett. 763(1), 7 (2013) https://arxiv.org/abs/1211.6804 [astro-ph.CO]. https://doi.org/10.1088/2041-8205/763/1/L7
31. Koekemoer, A.M., et al.: The 2012 Hubble Ultra Deep Field (UDF12): Observational Overview. Astrophys. J. Suppl. Ser. 209(1), 3 (2013) https://arxiv.org/abs/1212.1448 [astro-ph.CO]. https://doi.org/10.1088/0067-0049/209/1/3
32. Brammer, G.B., et al.: A Tentative Detection of an Emission Line at 1.6 µm for the z ~12 Candidate UDFj-39546284. Astrophys. J. Lett. 765(1), 2 (2013) https://arxiv.org/abs/1301.0317 [astro-ph.CO]. https://doi.org/10.1088/2041-8205/765/1/L2
33. Senchyna, P., et al.: Ultraviolet spectra of extreme nearby star-forming regions - approaching a local reference sample for JWST. Mon. Not. R. Astron. Soc. 472(3), 2608–2632 (2017) https://arxiv.org/abs/1706.00881 [astro-ph.GA]. https://doi.org/10.1093/mnras/stx2059
34. Senchyna, P., et al.: Extremely metal-poor galaxies with HST/-COS: laboratories for models of low-metallicity massive stars and high-redshift galaxies. Mon. Not. R. Astron. Soc. 488(3), 3492–3506 (2019) https://arxiv.org/abs/1904.01615 [astro-ph.GA]. https://doi.org/10.1093/mnras/stz1907
35. Nanayakkara, T., et al.: Exploring He II λ1640 emission line properties at z ~2-4. Astron. Astrophys. 624, 89 (2019) https://arxiv.org/abs/1902.05960 [astro-ph.GA]. https://doi.org/10.1051/0004-6361/201834565
36. Reddy, N.A. et al. The MOSDEF Survey: Significant Evolution in the Rest-frame Optical Emission Line Equivalent Widths of Star-forming Galaxies at z = 1.4–3.8 The Astrophys. J. 869, 92 (2018) https://iopscience.iop.org/article/10.3847/1538-4357/aaed1e
37. Chevallard, J., Charlot, S.: Modelling and interpreting spectral energy distributions of galaxies with BEAGLE. Mon. Not. R. Astron. Soc. 462(2), 1415–1443 (2016) https://arxiv.org/abs/1603.03037 [astro-ph.GA]. https://doi.org/10.1093/mnras/stw1756
38. Hirashita, H., Il'in, V.B., Pagani, L., Lefèvre, C.: Evolution of dust porosity through coagulation and shattering in the interstellar medium. Mon. Not. R. Astron. Soc. 502(1), 15–31 (2021) https://arxiv.org/abs/2101.02365 [astro-ph.GA]. https://doi.org/10.1093/mnras/staa4018
39. Chevallard, J., et al.: Physical properties and H-ionizing-photon production rates of extreme nearby star-forming regions. Mon. Not. R. Astron. Soc. 479(3), 3264–3273 (2018) https://arxiv.org/abs/1709.03503 [astro-ph.GA]. https://doi.org/10.1093/mnras/sty1461
40. Cullen, F., et al.: The ultraviolet continuum slopes (β) of galaxies at z≃8–15 from JWST and ground-based near-infrared imaging. Mon. Not. R. Astron. Soc. 520(1), 14-23 (2023) https://arxiv.org/abs/2208.04914 [astro-ph.GA]
41. Topping, M.W., et al.: Searching for Extremely Blue UV Continuum Slopes at z = 7 – 11 in JWST/NIRCam Imaging: Implications for Stellar Metallicity and Ionizing Photon Escape in Early Galaxies. The Astrophys. J. 941(2), 153 (2022) https://arxiv.org/abs/2208.01610 [astro-ph.GA]

42. Bouwens, R.J., et al.: Very Blue UV-Continuum Slope β of Low Luminosity z ~7 Galaxies from WFC3/IR: Evidence for Extremely Low Metallicities? Astrophys. J. Lett. 708(2), 69–73 (2010) https://arxiv.org/abs/0910.0001 [astro-ph.CO]. https://doi.org/10.1088/2041-8205/708/2/L69
43. Endsley, R., et al.: A JWST/NIRCam Study of Key Contributors to Reionization: The Star-forming and Ionizing Properties of UV-faint z ~ 7 – 8 Galaxies. arXiv e-prints, 2208–14999 (2022) https://arxiv.org/abs/2208.14999 [astro-ph.GA]
44. Greig, B., et al.: IGM damping wing constraints on reionization from covariance reconstruction of two z ≳ 7 QSOs. Mon. Not. R. Astron. Soc. 512(4), 5390–5403 (2022) https://arxiv.org/abs/2112.04091 [astro-ph.CO]. https://doi.org/10.1093/mnras/stac825
45. Davies, F.B., et al.: Predicting Quasar Continua near Lyα with Principal Component Analysis. Astrophys. J. 864(2), 143 (2018) https://arxiv.org/abs/1801.07679 [astro-ph.GA]. https://doi.org/10.3847/1538-4357/aad7f8
46. Ferruit, P., et al.: The Near-Infrared Spectrograph (NIRSpec) on the James Webb Space Telescope. II. Multi-object spectroscopy (MOS). Astron. Astrophys. 661, 81 (2022) https://arxiv.org/abs/2202.03306 [astro-ph.IM]. https://doi.org/10.1051/0004-6361/202142673
47. Birkmann, S.M., et al.: Wavelength calibration of the JWST near-infrared spectrograph (NIRSpec). Proc. SPIE 8150, 81500 (2011). https://doi.org/10.1117/12.893896
48. Böker, T., et al.: The spectro-photometric calibration of the JWST NIRSpec instrument. Proc. SPIE 8442, 84423 (2012). https://doi.org/10.1117/12.925369
49. Giardino, G., et al.: The Impact of Cosmic Rays on the Sensitivity of JWST/NIRSpec. Publ. Astron. Soc. Pac. 131(1003), 094503 (2019) https://arxiv.org/abs/1907.04051 [astro-ph.IM]. https://doi.org/10.1088/1538-3873/ab2fd6
50. Kriek, M., et al.: Direct Measurements of the Stellar Continua and Balmer/4000 Å Breaks of Red z > 2 Galaxies: Redshifts and Improved Constraints on Stellar Populations1,. Astrophys. J. 645(1), 44–54 (2006) https://arxiv.org/abs/astro-ph/0603063 [astro-ph]. https://doi.org/10.1086/504103
51. Calzetti, D., Kinney, A.L., Storchi-Bergmann, T.: Dust Extinction of the Stellar Continua in Starburst Galaxies: The Ultraviolet and Optical Extinction Law. Astrophys. J. 429, 582 (1994). https://doi.org/10.1086/174346
52. Chevallard, J., et al.: Simulating and interpreting deep observations in the Hubble Ultra Deep Field with the JWST/NIRSpec low-resolution 'prism'. Mon. Not. R. Astron. Soc. 483(2), 2621–2640 (2019) https://arxiv.org/abs/1711.07481 [astro-ph.GA]. https://doi.org/10.1093/mnras/sty2426
53. Bruzual, G., Charlot, S.: Stellar population synthesis at the resolution of 2003. Mon. Not. R. Astron. Soc. 344(4), 1000–1028 (2003) https://arxiv.org/abs/astro-ph/0309134 [astro-ph]. https://doi.org/10.1046/j.1365-8711.2003.06897.x
54. Vidal-García, A., Charlot, S., Bruzual, G., Hubeny, I.: Modelling ultraviolet-line diagnostics of stars, the ionized and the neutral interstellar medium in star-forming galaxies. Mon. Not. R. Astron. Soc. 470(3), 3532–3556 (2017) https://arxiv.org/abs/1705.10320 [astro-ph.GA]. https://doi.org/10.1093/mnras/stx1324
55. Gutkin, J., Charlot, S., Bruzual, G.: Modelling the nebular emission from primeval to present-day star-forming galaxies. Mon. Not. R. Astron. Soc. 462(2), 1757–1774 (2016) https://arxiv.org/abs/1607.06086 [astro-ph.GA]. https://doi.org/10.1093/mnras/stw1716
56. Chabrier, G.: Galactic Stellar and Substellar Initial Mass Function. Publ. Astron. Soc. Pac. 115(809), 763–795 (2003) https://arxiv.org/abs/astro-ph/0304382 [astro-ph]. https://doi.org/10.1086/376392
57. Charlot, S., Fall, S.M.: A Simple Model for the Absorption of Starlight by Dust in Galaxies. Astrophys. J. 539(2), 718–731 (2000) https://arxiv.org/abs/astro-ph/0003128 [astro-ph]. https://doi.org/10.1086/309250
58. Carton, D., et al.: Inferring gas-phase metallicity gradients of galaxies at the seeing limit: a forward modelling approach. Mon. Not. R. Astron. Soc. 468(2), 2140–2163 (2017) https://arxiv.org/abs/1703.01090 [astro-ph.GA]. https://doi.org/10.1093/mnras/stx545
59. Heckman, T.M., et al.: Extreme Feedback and the Epoch of Reionization: Clues in the Local Universe. Astrophys. J. 730(1), 5 (2011) https://arxiv.org/abs/1101.4219 [astro-ph.CO]. https://doi.org/10.1088/0004-637X/730/1/5
60. Vidal-García, A. et al.: BEAGLE-AGN I: Simultaneous constraints on the properties of gas in star-forming and AGN narrow-line regions in galaxies arXiv e-prints, 2211–13648 (2022) https://arxiv.org/abs/2211.13648 [astro-ph.GA]
61. Miralda-Escudé, J.: Reionization of the Intergalactic Medium and the Damping Wing of the Gunn-Peterson Trough. Astrophys. J. 501(1), 15–22 (1998) https://arxiv.org/abs/astro-ph/9708253 [astro-ph]. https://doi.org/10.1086/305799
62. Gunn, J.E., Peterson, B.A.: On the Density of Neutral Hydrogen in Inter-galactic Space. Astrophys. J. 142, 1633–1636 (1965). https://doi.org/10.1086/148444
63. Planck Collaboration, et al.: Planck 2015 results. XIII. Cosmological parameters. Astron. Astrophys. 594, 13 (2016) https://arxiv.org/abs/1502.01589 [astro-ph.CO]. https://doi.org/10.1051/0004-6361/201525830

64. McQuinn, M., Lidz, A., Zaldarriaga, M., Hernquist, L., Dutta, S.: Probing the neutral fraction of the IGM with GRBs during the epoch of reioniza- tion. Mon. Not. R. Astron. Soc. 388(3), 1101–1110 (2008) https://arxiv.org/abs/0710.1018 [astro-ph]. https://doi.org/10.1111/j.1365-2966.2008.13271.x